\documentstyle[preprint,aps]{revtex}

\begin{document}

\date{February 1, 1995}

\title{Normalization Sum Rule and Spontaneous
Breaking of U(N) Invariance  in Random Matrix Ensembles.}
\author{C.~ M.~Canali$^1$, and V.~E.~Kravtsov$^{1,2}$}
\address{ $^1$ International Centre for Theoretical Physics, 34100 Trieste,
Italy \\
$^2$Institute of Spectroscopy, Russian
 Academy of
Sciences,
142092 Troitsk, Moscow r-n, Russia}

\maketitle

\begin{abstract}
%
It is shown that the two-level correlation function $R(s,s')$ in the
invariant random matrix
ensembles (RME) with soft confinement exhibits a "ghost peak" at $s\approx
-s'$. This lifts the sum rule prohibition
for the level number variance to have a Poisson-like term ${\rm var}(n)=\eta n$
that is typical of
RME with broken $U(N)$ symmetry. Thus we conclude that the $U(N)$
invariance is broken spontaneously in the RME with soft confinement,
$\eta$ playing the role of an order-parameter.
\end{abstract}


The statistical description of complex systems by ensembles of random
matrices turned out to be a powerful general approach
that was successively applied to a great variety of systems in different
fields from nuclear physics [1] to mesoscopics [2] and quantum chaos [3].

The classical random matrix theory (RMT) by Wigner, Dyson and Mehta [1]
describes the statistics of eigenvalues for a Gaussian ensemble of random
Hermitian $N\times N$-matrices ${\bf H}$ with the probability distribution
$P({\bf H})\propto \exp[-Tr {\bf H}^2]$. By definition, the statistical
properties of this ensemble are invariant under unitary transformations
$U(N)$
of matrices ${\bf H}$ and thus there is no basis preference in the RMT. This
means that the classical RMT can be applied only to quantum systems where
all (normalized) linear combinations of eigenfunctions have similar
properties.

For disordered electronic systems, it implies that all
eigenstates must be extended. In other words, the classical RMT is
applicable only for describing the energy level statistics in the metal
phase [4,5] that exists in the dimensionality $d>2$ at
a relatively weak disorder.

With disorder increasing the system goes through the Anderson transition to
an insulating phase in which all eigenstates are localized. The level
statistics in this phase obviously cannot be described by the
$U(N)$-invariant RME, since one can construct an extended state by a linear
combination of localized states randomly positioned
throughout the sample. Thus the proper probability
distribution $P({\bf H})$ must contain a basis preference in order to
exclude unitary transformations which would lead to formation of such
extended states.

The ensemble of random banded matrices (RBME) [6,7] is an
example of such a non-invariant RMT. It describes properties
of systems belonging to so-called quasi-1d universality class which
includes quasi-1d disordered electronic systems with localization [6] and
certain quantum chaotic systems [7].
The corresponding eigenvalue statistics
are Poissonic in
the $N\rightarrow\infty$ limit
(at a fixed bandwidth $b$) and reduce to the Wigner-Dyson form at
 $b\rightarrow\infty$.
Thus changing the parameter $b/N$, one can describe
the
crossover from Wigner-Dyson to Poisson level statistics which occurs in
quasi-1d disordered systems with increasing the ratio $L/\xi$ of the sample
length $L$ and the localization radius $\xi$.

While the localization in quasi-1d systems seems to be well described in
terms of the RBME, the problem of the random matrix description
of the critical region near the Anderson transition and the Anderson
insulator phase
for $d>2$ remains open. The recent works
[8-10] where the existence of the universal critical level statistics
has been demonstrated, resulted in an intensive search for
the proper RME description.

In this connection, two different generalizations of the
classical RMT have been recently proposed [11,12]. The generalized
RME studied in Ref.[11] was obtained from the Gaussian
invariant ensemble by introducing a symmetry-breaking term:
\begin{equation}
\label{ShE}
P({\bf H})\propto e^{- Tr {\bf H}^2}\,e^{-h^2 N^2 Tr([{\bf \Lambda},{\bf
H}][{\bf \Lambda},{\bf H}]^{\dagger})}.
\end{equation}
The $h$-dependent term breaks the $U(N)$ invariance and tends to align
${\bf H}$ with a symmetry breaking unitary matrix ${\bf \Lambda}$ thus
setting the basis
preference. It turned out [11] that even after averaging over ${\bf
\Lambda}$ the
resulting ensemble leads to the eigenvalue statistics that deviate from
the Wigner-Dyson form. The difference between the Wigner-Dyson statistics
that correspond to $h=0$ and the level statistics for {\it any non zero} $h$
turns out to be dramatic in the thermodynamic (TD) limit
$N\rightarrow\infty$.
Namely, for $h\neq 0$ the variance ${\rm var}(n)$ of the number of levels in an
energy window that
contains $n$ levels on the average, grows linearly with $n$ at $n\gg 1$:
\begin{equation}
\label{var}
{\rm var}(n)=\langle (\delta n)^2\rangle=\eta(h) \,n\sim h\,n, \;\;\;\eta(0)=0.
\end{equation}
For the classical RMT [1], ${\rm var}(n)\propto \ln n$ that is negligible
as compared to Eq.(\ref{var}) for  {\it any} nonzero $0<\eta(h)<1$ in
the limit $n\rightarrow\infty$.
The Poisson-like behavior described by Eq.(\ref{var}) is valid also for
RBME in the TD limit.

In contrast to Eq.(\ref{ShE}), the probability distribution suggested in
Ref.[12] is explicitly $U(N)$-invariant:
\begin{equation}
\label{MutE}
P({\bf H})\propto e^{-Tr V({\bf H}) }.
\end{equation}
The only singularity in this model is that the
(even in $E$) "confining potential" $V(E)$
grows extremely slowly:
\begin{equation}
\label{V}
V(E)=\mbox{$\frac{A}{2}$} \ln^2 |E|,\;\;\;\;\;\; |E|\rightarrow\infty.
\end{equation}
However, the Poisson-like behavior, Eq.(\ref{var}), turns out to be
valid  for this model too, provided
that the energy window does not contain the origin $E=0$.

It should be stressed [13] that for steeper confining potentials, $V(E)=
|E|^{\alpha}$, no deviation from the
Wigner-Dyson statistics
was observed in the bulk of the spectrum, so that $\eta_{\alpha}=0$ for
{\it all} $\alpha>0$.
What happens with the invariant RME
at the transition from a power-law to logarithmic confinement, looks like
the spontaneous breaking of the $U(N)$-symmetry.

In this Letter we present both analytical and numerical arguments
showing this novel phenomenon to exist. We show the parameter $\eta$
to play the role of an order-parameter
and clarify its connection with the breaking of the
normalization sum rule in the TD limit.

The reason why $\eta=0$ for a wide class of invariant RME
is connected with the normalization sum rule:
\begin{equation}
\label{SR}
\int_{-\infty}^{+\infty} Y^{(N)}_{2}(s,s')ds'=1.
\end{equation}
Here the cluster function $Y_{2}(s,s')=\delta(s-s')-R(s,s')$ is
related to the two-level correlation function $R(s,s')$:
\begin{equation}
\label{TL}
R^{(N)}(s,s')=\frac{\langle
\rho(E_{s})\rho(E_{s'})\rangle}{\langle\rho(E_{s})\rangle\,\langle\rho(E_{s'})
\rangle}-1.
\end{equation}
where $\rho(E)=Tr\{\delta(E-{\bf H})\}$ is the level
density, $\langle...\rangle$ denotes ensemble averaging with the
probability distribution $P({\bf H})$, and
the new variable $s(E)$ is chosen so that the mean level density in this
variable, $\langle\tilde{\rho}(s)\rangle=\langle\rho(E_{s})\rangle
dE_{s}/ds=1$ for $-N/2<s<N/2$.

For $N$ finite,
Eq.(\ref{SR}) is an exact property of any RME
that follows simply from
the normalization condition $\int
\rho(E) dE=N$. However, the sum rule may be violated [14] in the
limit $N\rightarrow\infty$,
since after the integration in Eq.(\ref{SR}) both terms in Eq.(\ref{TL})
result in the
divergent constants $N$ to be subtracted from each other. Below we will
assume this limit to be taken.

The general relationship between the cluster function $Y_{2}(s,s')$ and the
coefficient
$\eta=\lim_{n\rightarrow\infty}\{d [{\rm var}(n)]/dn\}$ in the level number
variance, Eq(\ref{var}),
for an energy window centered at a point $E_{0}=E(s_{0})$, reads:
\begin{equation}
\label{dvdn}
\eta=\lim_{n\rightarrow\infty}\left[1-\int_{a_{-}}^{a_{+}}[Y^{\infty}_{2}
(a_{+},s')+
Y^{\infty}_{2}(a_{-},s')]\,ds'\right],
\end{equation}
where $a_{\pm}=s_{0}\pm n/2$.
In the case where the cluster function in the TD limit
$Y^{\infty}_{2}(s,s')=Y^{\infty}_{2}(s-s')$ is translationally
invariant, the relationship reduces to:
\begin{equation}
\label{dvdnTD}
\eta=1- \int_{-\infty}^{+\infty} Y^{\infty}_{2}(s)\,ds.
\end{equation}
Now comparing Eq.(\ref{SR}) and Eq.(\ref{dvdnTD}) we can conclude that the
parameter $\eta$ is equal to the deficiency of the sum rule
in the $N\rightarrow\infty$ limit [9].

Whether the sum rule is broken or not depends on the behavior of the
effective level interaction at large distances. Quite generally, the
joint probability distribution $P[\{x_{i}\}]$
of eigenvalues $x_{i}$ of a matrix ${\bf H}$ can be
represented in the form similar to the Gibbs distribution
$P[\{ x_{i}\}]=\exp[-\beta {\cal H}]$   for the one-dimensional plasma of
classical particles described by the potential energy functional
${\cal H}[\{x_{i}\}]= \sum_{i}V(x_{i})+W[\{x_{i} \}]$. For the unitary
ensembles considered in this Letter, the effective temperature $\beta=2$. In
case of the
invariant
ensembles given by Eq.(\ref{MutE}),
the many-body interaction  $W[\{x_{i} \}]$
reduces to
the pair-wise logarithmic effective level interaction [1]:
\begin{equation}
\label{log}
W[\{x_{i}\}]= -
\sum_{i>j}\ln|x_{i}-x_{j}|.
\end{equation}
For RME similar to that given by
Eq.(\ref{ShE}), the
corresponding effective level interaction that appears after averaging [11]
over the symmetry-breaking matrix ${\bf \Lambda}$,
contains all the
many-body terms and can be rewritten  in the form of a determinant
$Det_{ij}$, $i,j=1,...N$:
\begin{equation}
\label{det}
W[\{x_{i} \}]= - \mbox{$\frac{1}{2}$} \ln Det_{ij}[e^{-h^2 N^2
(x_{i}-x_{j})^2}].
\end{equation}
In both limiting cases, $|x_{i}-x_{j}|\rightarrow 0$ and
$|x_{i}-x_{j}|\rightarrow\infty$, the determinant in Eq.(\ref{det}) can be
approximated by a pair-wise interaction of the form:
\begin{equation}
\label{cut}
W[\{x_{i}\}]=
-\mbox{$\frac{1}{2}$} \sum_{i>j}
\ln(1-e^{-2h^2 N^2 (x_{i}-x_{j})^2})+const.
\end{equation}

The main difference between Eq.(\ref{log}) and Eq.(\ref{cut})
is that the symmetry-breaking field
$h\neq 0$ results in the cut-off of the effective level interaction
at sufficiently large distances [15]. This difference is
crucial for the fulfillment of the sum rule, Eq.(\ref{SR}), in the
TD limit.

It can be shown that the normalization sum
rule persists also in the TD limit if the effective
level interaction is long-range (non-integrable). We will present here
a simple proof of this statement based on the mean-field approach
[16] that
is valid exactly for the long-range interactions in the
$N\rightarrow\infty$ limit. Suppose that the effective level interaction
$f(x_{i}-x_{j})$ is pair-wise and long-range.
Then using the relationship $\delta
\langle\rho(E) \rangle/\delta V(E')= -\beta [\langle\rho(E)\rho(E')
\rangle-\langle\rho(E) \rangle\,\langle\rho(E') \rangle]$ and the
Dyson mean-field equation [1] for $\langle\rho(E)\rangle$
\begin{equation}
\label{MFE}
\int_{-\infty}^{+\infty}\langle\rho(E'')\rangle f(E''-E')+V(E')=const,
\end{equation}
one can
derive the integral equation for the two-level correlation function [16]:
\begin{equation}
\label{IE}
\int_{-\infty}^{+\infty} ds''
R^{\infty}(s,s'')\,f(E_{s''}-E_{s'})=\beta^{-1}\delta(s-s').
\end{equation}
Now integrate this equation over all $s$ changing the order of integration
in the left-hand side
and denote the integral $\int_{-\infty}^{+\infty} R^{\infty}(s,s'')ds=I$:
\begin{equation}
\label{div}
I\beta \int_{-\infty}^{+\infty}f(E_{s''}-E_{s})\,ds''=1.
\end{equation}
{}From the definition of the function $E_{s}$ just after Eq.(\ref{TL}) and
a physically obvious property $d\langle\rho(E) \rangle/d|E|<0$ that holds
for an even confinement potential $V(E)$, it follows that $E_{s}$ increases
linearly or faster with $s$. Therefore for any long-range interaction
with $\int f(E)dE$ divergent at infinity, the
integral in Eq.(\ref{div}) is also divergent. Thus we arrive at the
relationship:
\begin{equation}
\label{SRp}
I=1-
\int_{-\infty}^{+\infty}Y_{2}^{\infty}(s,s')ds'=0,
\end{equation}
which proves the validity of Eq.(\ref{SR}) in the limit
$N\rightarrow\infty$.
On the contrary, if the effective interaction $f(E-E')$ is cut at large
distances, the quantity $I$ in Eqs.(\ref{div}),(\ref{SRp}) must be finite
and thus the sum rule is violated.

We see that the symmetry-breaking term in Eq.(\ref{ShE}) leads to the
cut-off of the effective level interaction at large distances and thus to
the break-down of the normalization sum rule in the TD
limit.
This results in
a quasi-Poisson level number variance, Eq.(\ref{var}), governed by the
nonzero parameter $\eta$, Eq.(\ref{dvdnTD}).

The situation resembles the appearance of the long-range order in spin
systems in the external magnetic field ${\bf h}$ which breaks down the
rotational
invariance. In this case, the spin-spin correlator $\langle {\bf S}({\bf r})
{\bf S}({\bf r'}) \rangle\rightarrow m^2$ is constant at large distances,
the magnetization $|m|$ depending linearly on
$|{\bf h}|$. Since the spin-spin correlator is invariant under global
rotations of the spins, the order parameter, that is $|m|$, is unchanged
after averaging over the direction of the symmetry-breaking field ${\bf h}$.

In our problem, the quantity analogous to
$\langle {\bf S}({\bf r}){\bf S}({\bf r'}) \rangle$,
 which is invariant under global (independent on
$E$, $E'$) unitary transformations,
is the two-level correlation function $\langle
Tr\{\delta(E-{\bf H})\}Tr\{\delta(E'-{\bf H}) \} \rangle$. The
symmetry-breaking field $h{\bf \Lambda}$ plays the role of the magnetic field,
and
the parameter $\eta$ is
similar to the magnetization.
Like $|m|$, $\eta$ remains nonzero
after averaging over ${\bf \Lambda}$.

Now we return to consider the TD limit of the invariant
ensemble with soft
confinement, defined by Eqs.(\ref{MutE}),(\ref{V}). Since the effective
level interaction, Eq.(\ref{log}), is long-range, the sum rule,
Eq.(\ref{SR}), must hold for any
confining potential $V(E)$. However, for soft confinement
this does not necessarily lead to $\eta=0$.
The reason is the dramatic break-down of the translational
invariance
of the cluster function
$Y^{\infty}_{2}(s,s')=Y^{\infty}_{2}(s-s')+Y_{a}(s,s')$
which turns out to have an anomalous part $Y_{a}(s,s')$ that exhibits a
sharp  peak of width $A$ near $s=-s'$.

With the translational invariance broken in such a way, the level number
variance ${\rm var}(n)$ becomes
dependent on the position of the energy window $s_{0}$,
and can oscillate as a function of $n$. The parameter $\eta$ is given
by the general Eq.(\ref{dvdn}), where $n = 1, 2, ...$.
It is easy to show that the integral
in Eq.(\ref{dvdn}) reduces
approximately to
that in the sum rule, Eq.(\ref{SR}),
only provided that $n/2-|s_{0}|\gg A$, that is the
origin $E=s=0$ is far within the energy window. This means that the parameter
$\eta$ vanishes for a symmetric window $s_{0}=0$ and it is nonzero if
$|s_{0}|\gg n/2$.

Indeed, the Monte Carlo
simulations on the classical 1d plasma described by Eqs.(\ref{V}),(\ref{log})
show a dramatic difference in the level number variance in these two
cases.
A remarkable property of the model is that for a symmetric window, the
level number variance ${\rm var}(n)$ is {\it constant} for all integers
$n\gg1$. Thus the
"level rigidity" is even higher than that for the classical RMT where
${\rm var}(n)\propto\ln n$.

Though the anomalous part $Y_{a}(s,s')$ can be extracted from the exact
solution [12] for the particular confining potential obeying Eq.(\ref{V}),
its existence and importance was never mentioned
before. Below we present a simplified derivation of the cluster function
$Y_{2}^{\infty}(s,s')$ for small values of the parameter $q=e^{-\pi^2 A}$.

We start with the representation [1] of $Y_{2}^{\infty}(s,s')$ in terms
of the
orthonormal "wave functions" $\varphi_{i}(E)=p_{i}(E) exp(-V(E))$, where
$p_{i}(E)$ are orthogonal polynomials corresponding to the weight function
$e^{-2V(E)}$:
\begin{equation}
\label{Y}
Y_{2}^{\infty}(s,s')=
\frac{K^{2}(E_{s},E_{s'})}{K(E_{s},E_{s})K(E_{s'},E_{s'})}.
\end{equation}
Here the kernel $K(E,E')$ is given by [1]:
\begin{equation}
\label{K}
K(E,E')=\frac{1}{\pi C^2}
\,\frac{\varphi_{o}(E')\varphi_{e}(E)-\varphi_{e}(E')\varphi_{o}(E)}{E'-E},
\end{equation}
where $C$ is a constant and $\varphi_{o(e)}$ are $N\rightarrow\infty$ limits
of wave functions $\varphi_{N}(E)$ of the odd and even order, respectively.

For $q\ll 1$,
the wave functions
$\varphi_{o(e)}(E)$ have a "semi-classical" form that is a generalization
of the $N\rightarrow\infty$ limit of the oscillator wave functions for
a nonlinear $s(E)$:
\begin{equation}
\label{anz}
\varphi_{o}(E)=C\sin[\pi s(E)],\;\;\; \varphi_{e}(E)=C\cos[\pi s(E)].
\end{equation}
The arguments in $\sin$ and $\cos$ are chosen so that the density of zeros
$\rho_{0}=ds/dE$ coincides with the mean level density $\langle\rho(E)
\rangle$.

The function $s(E)$ can be easily found from  the known
solution of the Dyson  mean field equation (\ref{MFE}). Solving
Eq.(\ref{MFE}) with $f(E-E')=-\ln|E-E'|$ for the confining potential,
Eq.(\ref{V}), and taking the limit $N\rightarrow\infty$, we have:
\begin{equation}
\label{s(E)}
\langle\rho(E)\rangle=ds/dE=\mbox{$\frac{A}{2|E|}$},\;\;\;E_{s}=\lambda
e^{2|s|/A}\,{\rm sign}(s),
\end{equation}
where $\lambda$ is a constant of integration.

Now using Eqs.(\ref{Y})-(\ref{s(E)}) and the relationship [1] $\langle
\rho(E)\rangle=K(E,E)$ one can obtain an explicit form of the cluster
function $Y_{2}(s,s')$ for $q=e^{-\pi^2 A}\ll 1$, where $A$ is a
coefficient in Eq.(\ref{V}).
For $ss'>0$ we arrive at the translationally invariant expression [12]:
\begin{equation}
\label{Yn}
Y_{2}^{\infty}(s-s')=\frac{1}{\pi^2
A^2}\,\frac{\sin^2[\pi(s-s')]}{\sinh^2[(s-s')/A]},\;\;\;ss'>0.
\end{equation}
However, there are strong correlations for $ss'<0$ too:
\begin{equation}
\label{Ya}
Y_{a}(s,s')=\frac{1}{\pi^2
A^2}\,\frac{\sin^2[\pi(s-s')]}{\cosh^2[(s+s')/A]},\;\;\;ss'<0.
\end{equation}
This is just the anomalous part of the cluster function
discussed above. Its remarkable property is
a sharp peak at $s\approx -s'$ with a height that {\it does not}
decrease when $|s-s'|\sim 2|s|\rightarrow\infty$.

The existence of such a "ghost" peak in the two-level correlation function
is confirmed also by the Monte Carlo
simulations
on the effective plasma model with a soft
confinement [see Fig.1].
Such anomalous, infinitely long-range level correlations make it
possible  for the {\it invariant} random matrix ensemble defined by
Eqs.(\ref{MutE}), (\ref{V}) to exhibit the Poisson-like level number
variance ${\rm var}(n)=\eta n$ which is typical for RME with {\it broken}
$U(N)$ symmetry. Moreover, after the substitution $\pi^2 A=2/h\gg 1$, the
form of the "normal", translationally
invariant part of $Y^{\infty}_{2}(s,s')$ given by Eq.(\ref{Yn}) is identical
to that found in Ref.[11] for the matrix ensemble, Eq.(\ref{ShE}), with
symmetry breaking.

This is a crucial point in the chain of arguments in favour of the statement
that the $U(N)$ symmetry is {\it spontaneously broken} in the random
matrix ensemble, Eqs.(\ref{MutE}),(\ref{V}), with soft confinement.

The corresponding order parameter which  arises either due to the external
symmetry breaking field $h$ or spontaneously for a sufficiently soft
confinement, can be defined in the $N\rightarrow\infty$ as follows:
\begin{equation}
\label{op}
\eta=\lim_{s\rightarrow+\infty}\left[1-\int_{s'>0}Y^{\infty}_{2}(s,s')\,ds'
\right].
\end{equation}
It coincides with the coefficient in the quasi-Poisson term in the level
number variance, Eq.(\ref{var}) provided that the energy window does not
contain the origin.

In case of the spontaneously broken $U(N)$ symmetry, a complementary order
parameter can be introduced:
\begin{equation}
\label{cop}
\tilde{\eta}=\lim_{s\rightarrow+\infty}
\left[-\int_{s'<0}Y^{\infty}_{2}(s,s')\,ds'\right],
\end{equation}
which involves integration only over the anomalous part of the
cluster function. Because of the sum rule, Eq.(\ref{SR}), the sum
$\eta+\tilde{\eta}=0$ must be zero. Exploiting the spin analogy discussed
above, one can say that the spontaneous symmetry breaking is of the
AFM type, with $\eta$ and $\tilde{\eta}$ playing a
role of the sublattice magnetizations.

We conclude that the anomalous statistics of eigenvalues in RME
with soft
confinement is a result of the spontaneous
breaking of the $U(N)$ symmetry which exhibits itself in the
quasi-Poisson term in the level number variance and in the appearance of the
"ghost"  peak at $s=-s'$ in the two-level correlation function $R(s,s')$.
\acknowledgments

We are grateful to Yu~Lu, A.~Mirlin, E.~Tosatti and Mats~Wallin for
stimulating discussions.

\begin{figure}
\caption{MC results showing the existence of the "ghost"
correlation peak. The simulations are performed with $N=100$
 "particles" in the confining
potential $V(E)=\mbox{$\frac{A}{2}$}\ln^2(1+|E|)$ for $\beta=2$, and A=0.5.
The reference particle was mobile around $s=24$. The solid line in the
inset corresponds to Eq.(21) with oscillations being averaged out.}
\label{fig1}
\end{figure}
\end{document}